\documentclass[12pt]{iopart}
\usepackage{graphicx}
\usepackage{iopams}

\newcommand{\be}{\begin{equation}}              
\newcommand{\ee}[1]{\label{#1} \end{equation}}  

\newcommand{\bee}{\begin{eqnarray}}             
\newcommand{\eee}{\end{eqnarray}}               
\def\reff#1{(\ref{#1})}

\begin{document}

\title[Mixing-induced activity in open flows]{Mixing-induced activity in open flows \footnote{Paper published in Phys. Scr. {\bf{T132}}, 014035 (2008); \\[0.5mm] {\mbox{$\;\;\,$} see also Phys. Rev. Lett. {\bf{99}}, 184503 (2007)}}}

\author{Arthur V Straube$^{1,2}$ and Arkady Pikovsky$^2$}
\address{$^1$ Institut f\"ur Theoretische Physik, Technische Universit\"at Berlin, Hardenbergstr. 36, D-10623 Berlin, Germany}
\address{$^2$ Department of Physics and Astronomy, University of Potsdam, Karl-Liebknecht-Str. 24/25, D-14476 Potsdam-Golm, Germany}

\ead{arthur.straube@gmail.com}

\date{\today}

\begin{abstract}
We develop a theory describing how a convectively unstable
active field in an open flow is transformed into absolutely unstable by local mixing. Presenting
the mixing region as one with a locally enhanced effective
diffusion allows us to find the linear transition point to an
unstable global mode analytically. We derive the critical exponent
that characterizes weakly nonlinear regimes beyond the instability
threshold and compare it with numerical simulations of a full
two-dimensional flow problem. The obtained scaling law turns out
to be universal as it depends neither on geometry nor on the
nature of the mixing region.
\end{abstract}








\section{Introduction}

In many geophysical and laboratory flows active chemical and
biological processes occur. These active processes are crucially dependent on the nature of the flow. Especially important from the theoretical point of view becomes understanding the role of mixing. As we demonstrate below, even localized region of strong mixing introduced in the laminar flow is able to significantly change the overall picture of activity. Because the active processes accompanied by mixing are widespread, the outcomes of our analysis are highly relevant to practically important problems varying from chemical reactions in micro-mixers to plankton growth in ocean, see \cite{tel-etal-05} and references therein.

What turns out to be generic, the active processes quite often occur in an open rather than in a closed geometry. Here
the main issue is whether the throughflow is stronger or weaker
than the activity. One has to compare the velocity of the
throughflow with the velocity of the activity spreading due to
diffusion. If the throughflow is stronger, the activity is blown
away like a candle flame in a strong wind, in the opposite case
a sustained activity can be observed \cite{kuznetsov-etal-97,
mcgraw-menzinger-03, sandulescu-etal-06}. This simple picture is
valid, however, only for homogeneous media. Often additional
vortexes are superimposed on a constant throughflow, due to, e.g.,
wakes behind islands in ocean currents or mixing enforced by
revolving fan blades in laboratory experiments. We want to study
under which conditions such an additional kinematic mixing in a
strong open flow can lead to a transition to a sustained activity,
and to characterize this transition quantitatively.

Our main model is a reaction-advection-diffusion equation for the
dimensionless concentration of an active scalar field
$u(\mathbf{r},t)$
\begin{equation}
\frac{\partial u}{\partial t} +[\mathbf{V} +\mathbf{W}(\mathbf{r},t)]\cdot \nabla
u=D_0\nabla^2 u+au(1-u^p)\;. \label{eq1}
\end{equation}
Here $\mathbf{V}=(V,0,0)$ is a constant throughflow in
$x$-direction, $D_0$ is a molecular diffusion of the scalar field.
Activity is assumed to be of the simplest form: a linear growth
with rate $a$ with a saturation at $u=1$. The nonlinearity index
$p$ is typically integer (1 or 2) for chemical reactions, while
for biological populations a wide range of values of $p$ has been
recently reported~\cite{sibly-etal-05}. Mixing is described by a
spatially localized incompressible velocity field
$\mathbf{W}(\mathbf{r},t)$, its intensity is denoted as $W$. Note
that in the absence of fluid flow Eq.~\reff{eq1} is reduced to the
famous Kolmogorov-Petrovsky-Piskunov-Fisher (KPPF) model of an
active medium with diffusion (see, e.g., \cite{murray-93} for
original references, analysis, and applications of KPPF), while
for $a=0$ Eq.~\reff{eq1} describes a linear evolution of a passive
scalar in a flow. Model \reff{eq1} can be used for the description
of biological activity, where $u$ is, e.g., concentration of a
growing plankton, advected by oceanic currents
\cite{huisman-etal-06}; for a possible laboratory realization see
recent experiments  with an autocatalytic reaction in a Hele-Shaw
cell with a throughflow~\cite{leconte-etal-03}.

In the absence of flow, the diffusion causes the active state to
spread forming eventually a front with velocity
$V_f=2\sqrt{aD_0}$~\cite{van-saarloos-03}. Thus, for vanishing
mixing $W=0$, the activity is blown away provided $V>V_f$. For
this parameter range the instability in Eq.~\reff{eq1} is
convective and in the absence of external sources, no activity is
observed in the medium. A nontrivial state is, however, possible
if there is a localized source of the field $u$: then a growing
tail stretches from this source in the downstream direction, where
it eventually saturates at $u=1$. The linearized problem with a
point ($\delta$-function) source of intensity $\e$ can be readily
solved, yielding
\begin{equation}
 u(x)=\e(\tilde V)^{-1}\exp[xV/2D_0]\exp(-|x|\tilde V/2D_0)\;
\label{eq01}
\end{equation}
in  one-dimensional setups and
\begin{equation}
 u(\mathbf{r})=\e(2\pi D_0)^{-1}\exp[xV/2D_0]K_0(|\mathbf{r}|\tilde
 V/2D_0) \label{eq02}
\end{equation}
in two dimensions (where $K_0$ is the modified Bessel function,
$\tilde V=(V^2-V_f^2)^{1/2}$).
Note that in both solutions $u\sim\exp(\mu_{\mp} x)$ as
$x\to\pm\infty$, where $\mu_{\pm}=(2D_0)^{-1}(V\pm\tilde V)$.

\begin{figure}
\centering
\includegraphics[width=0.45\textwidth]{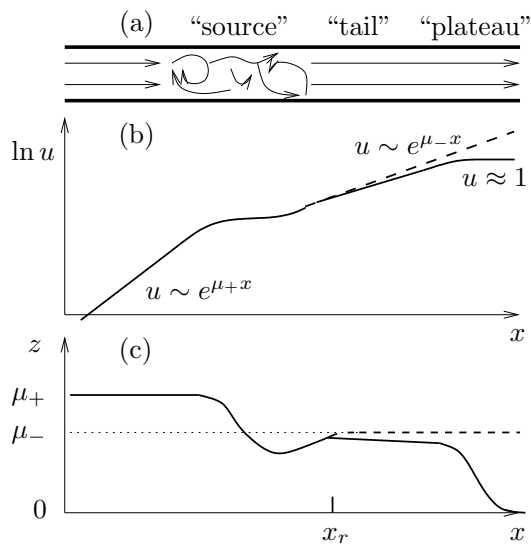}
\caption{(a): Quasi-one-dimensional open flow with a localized
mixing zone. Panels (b) and (c) illustrate the construction of the
nonlinear global mode, they show qualitative profiles $u(x)$ and
$z(x)=u^{-1}\frac{du}{dx}$ in the linear approximation at the
criticality (dashed line) and in nonlinear regime slightly beyond
criticality (full line).  In the latter case the profile is nearly
linear for $x<x_r$ but deviates due to nonlinear terms for
$x>x_r$, see discussion of Eq.~(\ref{eq10}). Regions ``source,''
``tail,'' and ``plateau'' are explained in the text.} \label{fig1}
\end{figure}

In this paper we demonstrate, that beyond some critical intensity
$W_{cr}$, a localized mixing $\mathbf{W}(\mathbf{r},t)$ turns the
convective instability locally into the absolute one, which
results in a stationary (in statistical sense) profile of $u$ (see
Fig.~\ref{fig1} for a sketch of the profile and Figs.~\ref{fig4}
and \ref{fig7} below for numerical examples). Beyond criticality
$W>W_{cr}$, the mixing region acts as an effective source of the
 field, in Fig.~\ref{fig1} this region is denoted as a
``source.'' A ``tail'' where the field grows exponentially as in
(\ref{eq01}), (\ref{eq02}) extends downstream of the source.
Further downstream stretches the ``plateau'' domain where $u=1$.
Our main quantitative result, obtained by matching solutions in
these three domains, is the critical exponent $\beta$ relating the
intensity of the effective source $\varepsilon_{eff}$ with the
mixing intensity: $\varepsilon_{eff}\sim (W-W_{cr})^\beta$.

To develop the theory we use the concept of global
modes~\cite{chomaz-etal-88, couairon-chomaz-97, tobias-etal-98}.
In this concept a self-sustained non-advected pattern arises due
to inhomogeneities of the system. Typically, one considers
inhomogeneities of the growth rate $a$: if $a=a(\mathbf{r})$ has a
hump where locally the front velocity is large $V_{f}^{loc}>V$,
then a global mode appears, located at this hump and downstream.
In this paper we are interested in another, mixing-based mechanism
of a global mode appearance. It can be easily understood if the
concept of effective diffusion (see, e.g.,~\cite{dimotakis-05}) is
used to describe the mixing term in \reff{eq1}. In this approach,
we phenomenologically introduce effective diffusivity
$D(\mathbf{r})=D_0+D_{mix}(\mathbf{r})$ that accounts for the
coarse grained mixing dynamics, and write instead of \reff{eq1} an
equation with a non-homogeneous diffusion
\begin{equation}
\frac{\partial u}{\partial t} +\mathbf{V}\cdot \nabla u=\nabla [D(\mathbf{r})\nabla
u]+au(1-u^p)\;. \label{eq3}
\end{equation}
A hump of diffusivity $D(\mathbf{r})$ leads to an increase of the
local front velocity $V_f$, and one expects that when the front
propagation prevails over the throughflow, a stationary global
mode can appear, producing a mixing-induced sustained structure.
We focus on a geometry shown in Fig.~\ref{fig1}(a): in a constant
open flow there is a localized region of strong mixing, which, as
we will see below, is not necessarily chaotic or turbulent. The
theory below will be developed for a one-dimensional case, which
is relevant, e.g., for flows in a micropipe; the results will be
supported by numerical simulations of two-dimensional flows. We
restrict ourselves to this case because of computational
simplicity, and also because two-dimensional flows are relevant
for many geophysical and laboratory (especially in microfluidics)
experimental situations.

\section{Linear stability analysis}

\subsection{Effective diffusion model}

We start with a linear analysis of a one-dimensional situation,
described by the linearized at $u=0$ Eq.~\reff{eq3}:
\begin{equation}
\frac{\partial u}{\partial t}+ V\frac{\partial u}{\partial
x}=\frac{\partial}{\partial x} \left[D(x)\frac{\partial
u}{\partial x}\right]+au\;. \label{eq4}
\end{equation}
We look for an exponentially growing in time solution and with an
ansatz $u(x,t)\sim \exp[\lambda t+\int^x z(\xi) d\xi]$ obtain
\begin{equation}
\frac{dz}{dx}=-z^2+\frac{V-\frac{d
D(x)}{dx}}{D(x)}z-\frac{a-\lambda}{D(x)}\;. \label{eq5}
\end{equation}
As $|x|\to \infty$ we have a homogeneous medium with $D=D_0$, here
the solution should tend to values
$z^0_{\pm}=(2D_0)^{-1}(V\pm\sqrt{V^2-4(a-\lambda)D_0})$ at which
the r.h.s. of \reff{eq5} vanishes. More precisely, as
$x\to-\infty$ we have $z\to z^0_+$ and as $x\to\infty$ we have
$z\to z^0_-$. With these two boundary conditions one easily finds
the solution of \reff{eq5} numerically, matching at $z=0$
integrations starting at large $|x|$ from the values $z^0_\pm$. In
a particular analytically solvable case of a piecewise-constant
diffusivity: $D=D_0$ for $|x|>l$ and $D=D_1>D_0$ for $|x|<l$, one
can perform the integration analytically and obtain the equation
for the growth rate $\lambda$:
\begin{equation}
l=\frac{2D_1}{\sqrt{4(a-\lambda)D_1-V^2}}\arctan\sqrt{\frac{V^2-4(a-\lambda)
D_0 } {4(a-\lambda)D_1-V^2 } }\;. \label{eq6}
\end{equation}
The value $\lambda=0$ corresponds to the onset of global
instability, in this case \reff{eq6} gives the relation between
the critical values $l_{cr}$ and $D_{1cr}$:
\begin{equation}
l_{cr}=\frac{2D_{1cr}}{\sqrt{4aD_{1cr}-V^2}}\arctan\sqrt{\frac{V^2-V_f^2
} {4aD_{1cr}-V^2 } }\;. \label{eq7}
\end{equation}
From \reff{eq7} it follows that $D_{1cr}\to\infty$ as $l_{cr}\to
l_{min}=\tilde V /(2a)$. In other words, there exists a minimal
size of the mixing region, so that for $l<l_{min}$ even a very
strong mixing, with a very large effective diffusion, cannot
create a global mode (the same is true for a smooth profile of
$D(x)$; note also that the size of the mixing region is not
limited from above). This is in contrast to the situation when the
global mode is induced by a local hump of the growth rate $a$
(cf.~\cite{dahmen-etal-00}): here one can obtain instability even
when $a(x)$ is highly localized (a delta-function), a global mode
then looks as in \reff{eq01}, \reff{eq02}.

\begin{figure}[htb]
\centering
\includegraphics[width=0.5\textwidth]{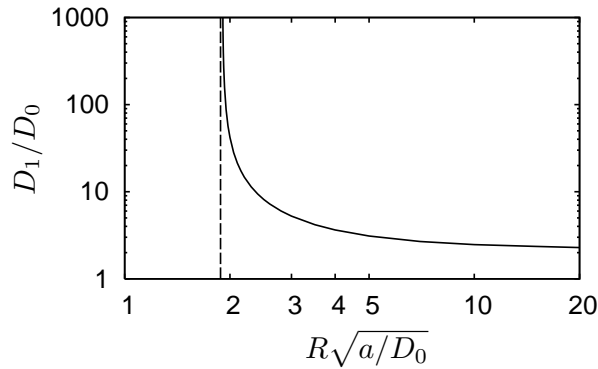}
\caption{The critical line for model~\reff{eq8}. Dashed line
denotes the minimal radius for the high-diffusive spot
$\sqrt{a/D_0}R_{min} \approx 1.905$.} \label{fig2}
\end{figure}

A similar analysis can be performed for a two-dimensional
inhomogeneous linear reaction-advection-diffusion equation \be
\frac{\partial u}{\partial t}+V\frac{\partial u}{\partial
x}=\nabla [D(r)\nabla u]+au\;, \ee{eq8} where we assume $D=D_1$
for $r<R$ and $D=D_0$ for $r>R$. Presenting the solutions in the
inner and outer domains as $u_i=\psi_i(r,\theta)\exp[(2D_1)^{-1}V
r\cos\theta+\lambda t]$ for $r<R$ and
$u_o=\psi_o(r,\theta)\exp[(2D_0)^{-1}V r\cos\theta+\lambda t]$ for
$r>R$, we obtain equations for $\psi_{i,o}$ whose solutions can be
written down as series in Bessel functions $J_m$ and $K_m$.
Matching these series at $r=R$ leads to rather cumbersome
expressions in terms of expansion coefficients. As a result, the
eigenvalue problem reduces to a matrix equation which we solved
numerically. In Fig.~\ref{fig2} we show the critical line
(corresponding to the condition $\lambda=0$) on the plane
$(R,D_1)$ for $V/V_f=3/2$. One can see, that similar to the
one-dimensional case, there exists a minimal radius of the higher diffusivity spot, so that for $R<R_{min}$ the global mode cannot arise.

\subsection{Non-uniform flow}

Now we discuss the linear stability not in the framework of the
effective diffusion model~\reff{eq3}, but in the full
reaction-advection-diffusion problem as in Eq.~\reff{eq1}. After
the linearization we arrive at a linear stability problem
\begin{equation}
\frac{\partial u}{\partial t} +[\mathbf{V} +\mathbf{W}(\mathbf{r},t)]\cdot \nabla
u=D_0\nabla^2 u+au\;, \label{eq801}
\end{equation}
which is non-stationary if the velocity field $\mathbf{W}$ is time
dependent. For a time-independent $\mathbf{W}$ the stability is
defined by the growth rate $\lambda$ as before. For time dependent
fields $\mathbf{W}$, the proper way to determine the stability is
to calculate the largest Lyapunov exponent (LE) $\Lambda=\langle
\frac{d}{dt}\ln||u||\rangle$. This can be done numerically, as
described in \cite{pikovsky-popovych-03}. Noteworthy, in this
consideration we are not restricted to a deterministic flow, as
the LE can be calculated also for a randomly or chaotically
time-dependent field $\mathbf{W}(\mathbf{r},t)$.

\begin{figure}[!htb]
\centering
\includegraphics[width=0.45\textwidth]{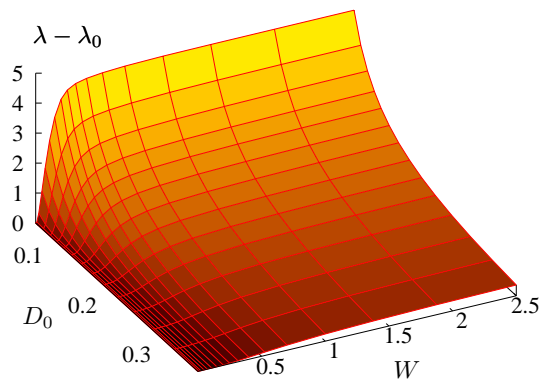}
\caption{Growth rate characterizing stability of the global mode
mixed by the flow with a stationary vortex \reff{streamf_1d}.} \label{fig3}
\end{figure}

We first calculate the growth rate $\lambda$ for a linearized
quasi-one-dimensional reaction-advection-diffusion equation
\reff{eq801} subject to a constant open flow with a superimposed
stationary vortex, described by a stream function
\begin{equation}
\Psi_1(x,y)=V y+W [\cos(2\pi y)-1]\exp(-x^2 b^{-2})\;.
\label{streamf_1d}
\end{equation}
We fix parameters $V=1$, $b=1$ and evaluate the growth rate for
different molecular diffusion constants $D_0$ and vortex
intensities $W$ (Fig.~\ref{fig3}). Note that the parameter $a$
simply shifts $\lambda$, therefore we plot $\lambda-\lambda_0$,
where $\lambda_0=a-V^2/(4D_0)$ is the growth rate for a non-mixed
flow with $W=0$. We see that the mixing-induced enhancement of
field growth is mostly pronounced for small diffusion and
saturates at $W\approx 0.5$.

\begin{figure}[!htb]
\centering
\includegraphics[width=0.55\textwidth]{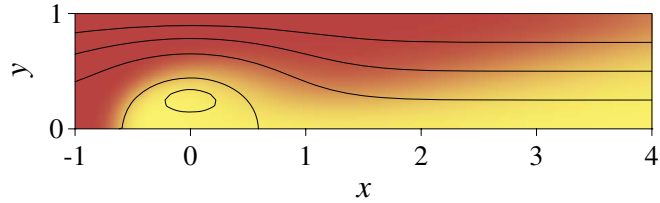}
\caption{An example of the global mode in the
quasi-one-dimensional flow \reff{streamf_1d} with $V=1$, $W=0.45$,
evaluated for $b=1$, $D_0=0.01$, $a=1.5$, $p=1$. Higher and lower
values of the active field concentration are plotted with the
lighter and darker colours, respectively.} \label{fig4}
\end{figure}

The role of a mixing vortex can be understood in the following
way. In the case of a passive advection-diffusion process and for a relatively small
molecular diffusivity, the field is trapped by the vortex.
In the limit of vanishing diffusivity the field $u$ outside the
vortex is blown away, whereas everything inside the vortex is
trapped and cannot escape for an infinitely long time, as e.g. in
\cite{lyubimov-etal-05}. As a result, a pattern in the form of a
cloud arises. Because of diffusion, such pattern is no longer stable and its concentration gradually decays. Hence, under an advection-diffusion process, the cloud exists only for a limited
time. However, if we now take into consideration activity, this
time is spent by the active field to grow, which is enough to compensate
the loss caused by diffusion. Thus, a self-sustained pattern is
born, Fig.~\ref{fig4}.

\begin{figure}
\centering
\includegraphics[width=0.45\textwidth]{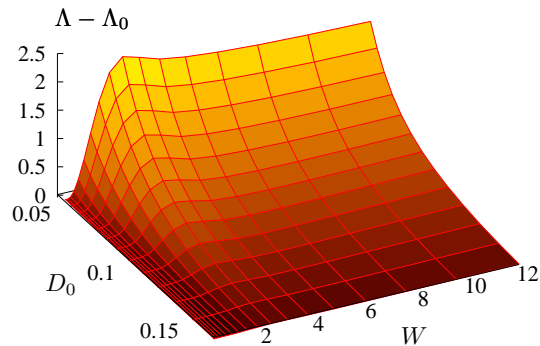}
\caption{Lyapunov exponent characterizing stability of the global
mode mixed by periodically blinking vortex \reff{eq20}.}
\label{fig5}
\end{figure}

Next we calculate the LE for a linearized two-dimensional
reaction-advection-diffusion equation \reff{eq801} with a constant
open flow and a superimposed oscillating vortex, described by the
stream function
\begin{equation}
\Psi_2(x,y,t)=V y+W\exp[-(x^2+y^2)R^{-2}]\cos(\omega t)\;.
\label{eq20}
\end{equation}
Here we fix $V=1$, $R=1$ and $\omega=2$ and calculate the LE
$\Lambda$, see Fig.~\ref{fig5}. As before, we plot
$\Lambda-\Lambda_0$, where $\Lambda_0=a-V^2/(4D_0)$ is the LE for
a non-mixed flow with $W=0$. Again, the mixing-induced enhancement
of field growth is mostly pronounced for small diffusion. In
contrast to the previous case the dependence on $W$ is
non-monotonic and has a maximum at $W\approx 3$. This is the
mixing strength at which a chaotic saddle~\cite{tel-etal-05} in
the Lagrangean particle trajectories appears. A further increase
of the vortex intensity does not lead, however, to a significant
growth of the LE.

\section{Nonlinear analysis. A universal scaling law}

Now we develop a nonlinear theory of the global mode. It is clear
that the nonlinear saturation stops the exponential growth of a
slightly supercritical linear mode and leads to a nonlinear
solution with a finite amplitude. Our aim is to describe the
dependence of this amplitude on the deviation from criticality.
First we notice, that the very notion of the amplitude is here
nontrivial. Indeed, the nonlinear solution looks as in
Fig.~\ref{fig1}(b) (cf. Fig.~\ref{fig7} below); it saturates to
$u=1$ in the downstream direction. However, outside the mixing
region the field looks like a solution caused by a localized field
source. Thus, we can take the effective intensity of this source
$\varepsilon_{eff}$, which is proportional to the characteristic
field amplitude in the mixing region $u(0)$ [see relations
(\ref{eq01}), (\ref{eq02})], as the order parameter of the
transition. The deviation from the criticality we will measure
with the growth rate $\lambda$, for which holds $\lambda\propto
W-W_{cr}$ in the full model \reff{eq1} or $\lambda\propto
D-D_{cr}$ in model \reff{eq3}.

We will consider the simplest possible setup, namely the nonlinear
modification of one-dimensional Eq.~\reff{eq4}:
\begin{equation}
\frac{\partial u}{\partial t}+V\frac{\partial u}{\partial
x}=\frac{\partial}{\partial x}\left[ D(x)\frac{\partial
u}{\partial x}\right]+au(1-u^p)\;. \label{eq81}
\end{equation}
We look for a stationary global mode $u(x)$, and rewrite this
equation as the system
\begin{eqnarray}
\frac{dz}{dx}&=-z^2+\frac{V-\frac{d D(x)}{dx}}{D}z-\frac{a}{D}+\frac{a}{D}u^{p}\;,\label{eq9-1}\\
\frac{du}{dx}&=zu\;.\label{eq9-2}
\end{eqnarray}
We consider this system separately in two spatial domains. The
first, linear region, includes the inflow and the mixing domains
(``source'' in Fig.~\ref{fig1}): $-\infty<x<x_r$, where the field
$u(x)$ remains small. In the second, outflow region
$x_r<x<\infty$, the field $u$ further grows (``tail'') and
nonlinearly saturates (``plateau''). In the linear region, because
of smallness of the field, we can neglect $u^p$ in \reff{eq9-1},
thus we obtain an equation similar to \reff{eq5}. The only
difference is that because we look for a stationary solution, in
\reff{eq9-1} the term $\sim \lambda$ is absent. Near the
criticality, where $\lambda$ is small, we can consider this term
as a perturbation, therefore the solution of \reff{eq9-1} in the
linear region is close to the solution of Eq.~\reff{eq5}; it has
the asymptotic $z\to \mu_+$ as $x\to -\infty$. Due to the
perturbation term $\propto \lambda$, at the right border of the
linear region $z$ deviates from $\mu_-$: the deviation $\mu_{-}-
z(x_r)$ is proportional to $\lambda$, and, thus, to $D-D_{cr}$. At
$x_r$ the field $u$ is small and $u(x_r)\propto u(0)$.

Next we consider full equations \reff{eq9-1}, \reff{eq9-2} in the
nonlinear region $x>x_r$. Here the solution should tend as
$x\to\infty$ to the saddle fixed point $u=1$, $z=0$. Thus,
starting integration from large values of $x$ in the negative
direction, we have to follow the stable manifold of this saddle
and match this solution at $x=x_r$ with the obtained above.
Because the value to be matched $z(x_r)$ is very close to $\mu_-$,
in the region where the solution $(z,u)$ approaches
$(z(x_r),u(x_r))$ we can write $z^2\approx\mu_-^2-2\mu_-\Delta z$
to obtain
\begin{equation}
\frac{d}{dx}\Delta z=(\mu_+-\mu_-)\Delta
z-\frac{a}{D}u^p(x_r)e^{p\mu_- (x-x_r)}\;. \label{eq10}
\end{equation}
Here, since $\Delta z=\mu_--z(x)$ is small, we have approximated
the solution of \reff{eq9-2} as $u\approx u(x_r)e^{\mu_-
(x-x_r)}$. Because linear inhomogeneous Eq.~\reff{eq10} is solved
in the \textit{negative} in $x$ direction, the solution follows
the \textit{slowest} exponent: $\Delta z\propto \exp[\gamma
(x-x_r)]$, where $\gamma=\min(\mu_+-\mu_-,p\mu_-)$.

At the criticality, the region of validity of the exponential
solution $\Delta z\propto \exp[\gamma (x-x_r)]$ becomes very
large. Thus it is dominant for small deviations from criticality
$\lambda$, therefore we can estimate the coordinate $x_s$  at
which the field $u$ saturates (i.e., we reach the state $u\approx
1$ and $z\approx 0$) from the relations above: from $-\mu_-\approx
(z(x_r)-\mu_-)e^{\gamma (x_s-x_r)}$ it follows $(x_s-x_r) \approx
-\gamma^{-1}\ln (\mu_- - z(x_r))$. Substituting this in the
expression for $u(x)$, we obtain $u(x_r)\propto(\mu_- -
z(x_r))^{\mu_- /\gamma}$. Now we take into account that $\mu_- -
z(x_r)\propto D-D_{cr}$, and,  because the evolution of $u$ in the
interval $0<x<x_r$ only weakly depends on the criticality,
$\varepsilon_{eff}= u(0)\propto u(x_r)$. The final expression for
the scaling law of the amplitude of the global mode thus reads
\begin{equation}
\varepsilon_{eff}  \propto \lambda^\beta \;, 
 \qquad \beta=\frac{\mu_-}{\gamma}=\max\left(\frac{\mu_-}{\mu_+-\mu_-},\frac{1}{p}\right)\;.\label{eq12}
\end{equation}
The critical index $\beta$ depends only on the nonlinearity index
$p$ and on the dimensionless velocity $v=V/V_f$:
\begin{equation}
\beta=\left\{
\begin{array}{ll}
p^{-1} \;\; & \mbox{if} \;\; v>\frac{2+p}{2\sqrt{1+p}}\;, \\
\frac{v-\sqrt{v^2-1}}{2\sqrt{v^2-1}} \;\; & \mbox{if} \;\;
1<v<\frac{2+p}{2\sqrt{1+p}}\;.
\end{array}
\right. \label{eq13}
\end{equation}
This main result of our letter can be physically interpreted as
follows. The exponent $\beta$ is determined solely by the
nonlinearity index $p$ if the throughflow velocity is much larger
than the front velocity ($v$ large). Here the field in the plateau
domain (see Fig.~\ref{fig1}) is effectively uncoupled from the
source, and the saturation of the instability is due to the local
nonlinearity at the source. For a small throughflow velocity ($v$
close to one) the plateau state interacts with the source via the
tail. Due to this ``remote control,'' the field at the source is
saturated more efficiently than due to nonlinearity, here the
exponent $\beta$ is determined solely by the form of the tail,
which depends on the velocities ratio $v$.

\begin{figure}
\centering
\includegraphics[width=0.45\textwidth]{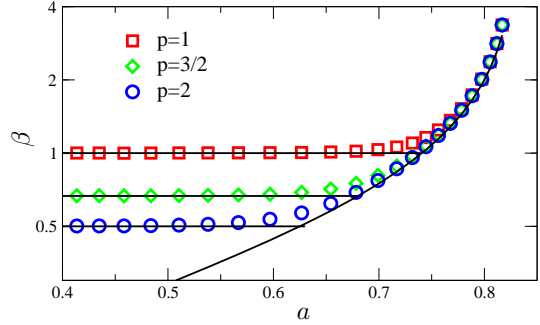}
\caption{The critical exponent $\beta$ calculated for the model
\reff{eq1} (symbols), compared with theoretical prediction
\reff{eq13} (lines). One can clearly see the crossover between two
regimes of the field saturation in dependence on parameter $a$,
the latter is related to $v$ in \reff{eq13}  via $v\sim
a^{-1/2}$.} \label{fig6}
\end{figure}

Below we check formula \reff{eq13} with direct numerical
simulations of model \reff{eq1}. A stationary vortex \reff{eq20}
with $\omega=0$ and $R=1$ was imposed on a constant flow with
$V=1$. Keeping the diffusion constant fixed $D_0=0.3$, for
different field growth rates $a$ we have found, from the
linearized equations, the critical vortex intensities $W_{cr}$ at
which the global mode becomes first unstable. Then we solved full
nonlinear equations close to criticality and found the exponent
$\beta$ according to \reff{eq12}. The stationary problem was
solved with a finite difference method in a domain $0\leq x\leq
60$, $0\leq y\leq 40$ with periodic boundary conditions in $y$ and
conditions $u(0)=0$, $\frac{\partial u}{\partial x}(60)=0$. The
results are presented in Fig.~\ref{fig6}, they are in very good
agreement with the theoretical prediction \reff{eq12},
\reff{eq13}. Figure~\ref{fig7} shows the example of the stationary
mode appearing beyond the instability threshold. A similar
analysis performed for the quasi-one-dimensional flow
\reff{streamf_1d} also provides very good agreement with formula
\reff{eq13}.
\begin{figure}[!htb]
\centering
\includegraphics[width=0.5\textwidth]{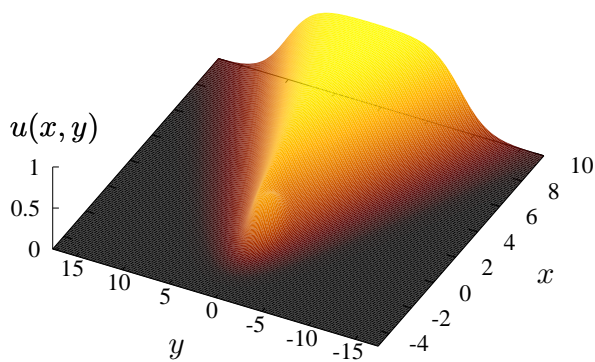}
\caption{The active field behind a vortex with $W=4$ placed at
$x=y=0$, for $V=1$, $a=0.5$, $D_0=0.3$, $p=2$. } \label{fig7}
\end{figure}

\section{Conclusions}

We have described the mixing-induced transition from a
convectively unstable active field in an open flow to a persistent
global mode. Our theoretical approach bases on the representation
of the mixing region as a domain with locally enhanced effective
diffusion. For a nonlinear regime close to criticality we have
derived the critical exponent $\beta$ \reff{eq13} that depends
only on two parameters of the system: the dimensionless flow
velocity $v$ normalized by that of the front, and the nonlinearity
index $p$. For large velocities the critical exponent depends only
on the system's nonlinearity, which means a local in space
saturation of the instability. For small velocities the exponent
is a function of velocity, here the growing downstream tail of the
active field imposes the saturation. Notably, this prediction of
the one-dimensional theory is in a good accordance with
two-dimensional calculations. The obtained results are independent of the geometry and the
nature of the mixing region and for these reasons are expected in different systems and at different scales. The generic nature of our findings indicates that turbulent mixing can play a key role in open flows that involve active chemical and biological processes.

\ack

We thank B~Eckhardt, U~Feudel, D~Goldobin, E~Knobloch, S~Kuznetsov, S~Shklyaev, and N Shnerb for fruitful discussions and DFG (SPP 1164 ``Nano- and Microfluidics,'' project STR 1021/1) for support.
%


\end{document}